\documentclass[%
 reprint,
superscriptaddress,
noeprint,
 amsmath,amssymb,
 aps,
]{revtex4-2}

\usepackage{graphicx}
\usepackage{dcolumn}
\usepackage{bm}
\usepackage{amsmath,amsfonts}
\usepackage{verbatim}
\usepackage[dvipsnames]{xcolor}

\usepackage[colorlinks,
bookmarks=true,
linkcolor=blue,
urlcolor=blue,
anchorcolor=black,
citecolor=blue
]{hyperref}
\usepackage{color}
\usepackage[normalem]{ulem}
\usepackage{xcolor}

\usepackage{hyperref}
\newcommand{\appref}{\hyperref[Sec:ind]{Appendix}}

\renewcommand{\sout}{\bgroup \color{red} \ULdepth=-.5ex \ULset}

\begin{document}

\preprint{APS/123-QED}

\title{Symmetry energy and neutron matter equation of state at $\rho_0/3$ from the electric dipole polarizability in $^{48}$Ca, $^{68}$Ni and $^{208}$Pb}

\author{Mengying Qiu}
\affiliation{Sino-French Institute of Nuclear Engineering and Technology, Sun Yat-Sen
University, Zhuhai 519082, China}

\author{Lie-Wen Chen}
\affiliation{State Key Laboratory of Dark Matter Physics, Key Laboratory for Particle Astrophysics and Cosmology (MOE), and Shanghai Key Laboratory for Particle
Physics and Cosmology, School of Physics and Astronomy, Shanghai Jiao Tong University, Shanghai, 200240, China}

\author{Zheng Zheng Li}
\affiliation{State Key Laboratory of Nuclear Physics and Technology, School of Physics, Peking University, Beijing 100871, China}
\affiliation{Frontier Science Center for Rare isotope, Lanzhou University, Lanzhou 730000, China}
\affiliation{School of Nuclear Science and Technology, Lanzhou University, Lanzhou 730000, China}

\author{Yi Fei  Niu}
\affiliation{Frontier Science Center for Rare isotope, Lanzhou University, Lanzhou 730000, China}
\affiliation{School of Nuclear Science and Technology, Lanzhou University, Lanzhou 730000, China}
\affiliation{Joint Department for Nuclear Physics, Lanzhou University and Institute of Modern Physics, CAS, Lanzhou 730000, China}
\affiliation{Department of Nuclear Physics, China Institute of Atomic Energy, Beijing, 102413, China}

\author{Zhen Zhang}
\email[Corresponding author:]{zhangzh275@mail.sysu.edu.cn}
\affiliation{Sino-French Institute of Nuclear Engineering and Technology, Sun Yat-Sen
University, Zhuhai 519082, China}

\date{\today}
\date{\today}

\begin{abstract}
Based on the quasiparticle random phase approximation implemented via the finite amplitude method, we employ a set of representative relativistic mean-field models to investigate the sensitivity of the inverse electric dipole polarizability $1/\alpha_{\mathrm{D}}$ in $^{48}\mathrm{Ca}$, $^{68}\mathrm{Ni}$, and $^{208}\mathrm{Pb}$ to the symmetry energy $E_{\rm{sym}}(\rho)$ and the neutron matter equation of state $E_{\rm{PNM}}(\rho)$ at a subsaturation density of $\rho = \rho_0/3$. Combined with predictions from nonrelativistic Skyrme energy density functionals (EDFs), our results reveal strong linear correlations between $1/\alpha_{\mathrm{D}}$ and both $E_{\rm{sym}}(\rho_0/3)$ and $E_{\rm{PNM}}(\rho_0/3)$. In particular, the $1/\alpha_{\mathrm{D}}$--$E_{\rm{PNM}}(\rho_0/3)$ correlation for $^{208}\mathrm{Pb}$ is found to be nearly model-independent. A Bayesian analysis of the measured values of $\alpha_{\rm{D}}$ in $^{48}\mathrm{Ca}$, $^{68}\mathrm{Ni}$, and $^{208}\mathrm{Pb}$ yields quantitative constraints of $E_{\mathrm{sym}}(\rho_0/3) = 17.8^{+1.1(1.8)}_{-0.9(1.6)}~\mathrm{MeV}$ and $E_{\mathrm{PNM}}(\rho_0/3) = 9.1^{+0.8(1.4)}_{-0.9(1.4)}~\mathrm{MeV}$ at the 68\% (90\%) confidence level, respectively. The extracted value of $E_{\mathrm{PNM}}(\rho_0/3)$ exceeds most predictions from microscopic many-body theories, suggesting a mild tension between nuclear EDF-based constraints derived from $\alpha_{\mathrm{D}}$ data and results from \textit{ab initio} calculations.

\end{abstract}

\maketitle


\section{\label{sec:level1}Introduction}

The nuclear symmetry energy $E_{\rm{sym}}(\rho)$, which characterizes the isospin dependence of nuclear equation of state (EOS), plays a critical role in both nuclear physics and astrophysics~\cite{Baran2005,Steiner:2004fi,Lattimer:2006xb, Li:2008gp, Oertel:2016bki,Roca-Maza:2018ujj}. Its density dependence has profound implications for various important phenomena, including the structure of neutron-rich nuclei, isospin dynamics in heavy-ion collisions, and the properties of neutron stars, neutron star mergers, and core-collapse supernovae. 
Closely related to $E_{\rm{sym}}(\rho)$ is  the EOS of pure neutron matter $E_{\rm{PNM}}(\rho)$, which is essential for describing neutron-rich systems. Knowledge of $E_{\rm{PNM}}(\rho)$ is of particular interest as it provides unique opportunities to test nuclear forces and many-body theories~\cite{Machleidt:2024bwl}. Consequently, a lot of work has been devoted to constrain the $E_{\rm{sym}}(\rho)$ or $E_{\rm{PNM}}(\rho)$ via various nuclear theories, experimental data and astrophysical observations.

Among various experimental observables for $E_{\rm{sym}}(\rho)$ and $E_{\rm{PNM}}(\rho)$, the electric dipole polarizability \(\alpha_{\rm D}\) is believed to be a clean and effective isovector probe. As early as 1944, Migdal pointed out its intimate connection to the symmetry energy entering in the semi-empirical mass formula~\cite{migdal1944quadrupole}. In the 1980s, the important role of surface symmetry energy in the evaluation of $\alpha_{\rm{D}}$ was noticed, and the experimental data on $\alpha_{\rm{D}}$ were used to constrain the surface symmetry energy~\cite{Lipparini:1981rew,Krivine:1982aok, MEYER1982269}. With advances in modern nuclear theories and experimental techniques, the $\alpha_{\rm{D}}$ has re-attracted a lot of attentions, especially since Ref.~\cite{Reinhard:2010wz}  proposed the $\alpha_{\rm{D}}$ as a sensitive probe of the neutron skin thickness 
and the density slope parameter of the symmetry energy. 
It was soon realized that the 
$\alpha_{\rm{D}}$ in $^{208}$Pb sensitively depends on both the magnitude and density slope of the symmetry energy at saturation density $\rho_0$, i.e., $E_{\rm{sym}}(\rho_0)$ and $L$. Moreover, the product $\alpha_{\rm{D}} \cdot E_{\mathrm{sym}}(\rho_0)$ has been shown to correlate even more strongly with $L$ as well as the neutron skin thickness~\cite{Roca-Maza:2013mla}.  The strong linear $\alpha_{\rm{D}} E_{\mathrm{sym}}(\rho_0)$--$L$ correlations have been further confirmed for other nuclei such as  $^{48}$Ca, $^{68}$Ni, and Sn isotopes~\cite{Roca-Maza:2015eza, Li:2021aij}.

While most studies have focused on the $\alpha_{\rm D}$--$L$  or $\alpha_{\rm{D}} E_{\mathrm{sym}}(\rho_0)$--$L$ correlation,   Ref.~\cite{Zhang:2015ava} has demonstrated that the $\alpha_{\mathrm{D}}$ of $^{208}\mathrm{Pb}$ is essentially determined by the symmetry energy or the EOS of neutron matter at a subsaturation density of $\rho = 0.05~\rm{fm}^{-3}\approx \rho_0/3$. The observed correlation of the form $1/\alpha_{\rm D}\propto E_{\rm{sym}}(\rho_0/3)$ naturally explains the simultaneous sensitivity of $\alpha_{\rm{D}}$  to both the $E_{\rm{sym}}(\rho_0)$ and $L$, and the proposed $1/\alpha_{\rm{D}}$--$E_{\rm{PNM}}(0.05~\rm{fm}^{-3})$ suggest  $\alpha_{\rm{D}}$ as a unique probe of neutron matter EOS.  While the established correlations in Ref.~\cite{Zhang:2015ava}  have been applied solely to $^{208}$Pb to extract constraints on $E_{\rm{sym}}(\rho)$ and $E_{\rm{PNM}}$ around $\rho_0/3$, it would be interesting to examine the correlations for other nuclei and to extract relevant information from more data from recent experiments. 

Experimentally, the $\alpha_{\rm{D}}$ was extracted from measurements of the photoabsorption or photo-neutron cross sections~\cite{Bohigas:1981fxj}. Recent high-precision experiments allow more accurate determination of the $\alpha_{\rm{D}}$. For example, proton inelastic scattering experiments at  forward angles performed at RCNP, Osaka, have recently measured the $\alpha_{\rm{D}}$ in  $^{40}$Ca~\cite{Fearick:2023lyz}, $^{48}$Ca~\cite{Birkhan:2016qkr}, $^{58}$Ni~\cite{Brandherm:2024rci}, $^{112,114,116,118,120,124}$Sn~\cite{Bassauer:2020iwp, Hashimoto:2015ema} and $^{208}$Pb~\cite{Tamii:2011pv}, while the $\alpha_{\rm{D}}$ in unstable nucleus $^{68}$Ni has been measured at GSI through Coulomb excitation  in inverse kinematics~\cite{Rossi:2013xha}. These experimental progress provides great opportunities to explore the EOS of asymmetric nuclear matter via nuclear electric dipole polarizabilities. 

In this work, based on the finding in Ref.~\cite{Zhang:2015ava}, we examine the dependence of  $\alpha_{\rm{D}}$  in neutron-rich 
doubly (semi-)magic nuclei $^{48}$Ca,  $^{68}$Ni and $^{208}$Pb 
on the symmetry energy and neutron matter EOS at $\rho_0/3$, i.e., $E_{\rm{sym}}(\rho_0/3)$ and $E_{\rm{PNM}}(\rho_0/3)$. The analysis is based on
a number of representative nuclear energy density functionals (EDFs), including three classes of relativistic mean-field (RMF) model, i.e., nonlinear meson-exchange, 
density-dependent meson-exchange  and density-dependent point-coupling models, as well as the nonrelativistic Skyrme EDFs. Based on predictions of these nuclear EDFs, the $E_{\rm{sym}}(\rho_0/3)$ and $E_{\rm{PNM}}(\rho_0/3)$ are constrained by Bayesian analysis of $\alpha_{\rm{D}}$ in  $^{48}$Ca,  $^{68}$Ni and $^{208}$Pb. 

The paper is organized as follows. In Sec.~\ref{Sec:form}, we introduce the theoretical framework, including the employed relativistic nuclear EDFs and the finite amplitude method used to compute the electric dipole polarizability. In Sec.~\ref{Sec:result}, results are presented for the correlations between the electric dipole polarizability and $E_{\rm{sym}}(\rho_0/3)$ and $E_{\rm{PNM}}(\rho_0/3)$, and the obtained constraints on $E_{\rm{sym}}(\rho_0/3)$ and $E_{\rm{PNM}}(\rho_0/3)$. A summary and conclusions are given in Sec.~\ref{Sec:conclusion}.

\section{Formalism\label{Sec:form}}
In this section, we present the theoretical formalism used to compute the electric dipole polarizability $\alpha_{\rm{D}}$. The calculations are performed using the quasiparticle finite amplitude method (QFAM) implemented within a set of representative relativistic mean-field models. The adopted relativistic nuclear EDFs are based on three types of effective Lagrangians, i.e., nonlinear meson-exchange (NL), density-dependent meson-exchange (DDME), and density-dependent point-coupling (DDPC) models. We first provide a brief introduction to the corresponding Lagrangian densities, followed by a description of the QFAM and the calculation of $\alpha_{\rm{D}}$.
\subsection{Nonlinear meson-exchange RMF model}
In the nonlinear (NL) RMF model, nucleons interact through the exchange of mesons, typically including the isoscalar-scalar $\sigma$, the isoscalar-vector $\omega$ meson and the isovector-vector $\rho$ mesons. The conventional NL effective Lagrangian density incorporates nonlinear self-interactions of the $\sigma$ and $\omega$ mesons, as well as a cross-coupling between $\rho$ and $\omega$ meson. It reads~\cite{Horowitz:2000xj} 
\begin{equation}\label{lagNL}
\begin{aligned}
\begin{aligned}
\mathcal{L}_{\mathrm{NL}}= & \bar{\psi}\left(i\gamma_\mu \partial^\mu -m\right) \psi-e \bar{\psi} \gamma_\mu \frac{1+\tau_3}{2} A^\mu \psi-\frac{1}{4} F^{\mu \nu} F_{\mu \nu} \\
& +g_\sigma \sigma \bar{\psi} \psi-g_\omega \omega_\mu \bar{\psi} \gamma^\mu \psi-g_\rho \vec{\rho}_\mu \bar{\psi} \gamma^\mu \vec{\tau} \psi \\
& +\frac{1}{2} \partial_\mu \sigma \partial^\mu \sigma-\frac{1}{2} m_\sigma^2 \sigma^2-\frac{1}{3} b_\sigma m\left(g_\sigma \sigma\right)^3-\frac{1}{4} c_\sigma\left(g_\sigma \sigma\right)^4 \\
& -\frac{1}{4} \omega_{\mu \nu} \omega^{\mu \nu}+\frac{1}{2} m_\omega^2 \omega_\mu \omega^\mu+\frac{1}{4} c_\omega\left(g_\omega^2 \omega_\mu \omega^\mu\right)^2 \\
& -\frac{1}{4} \vec{\rho}_{\mu \nu} \vec{\rho}^{\mu \nu}+\frac{1}{2} m_\rho^2 \vec{\rho}_\mu \vec{\rho}^\mu+\frac{1}{2} \Lambda_V\left(g_\rho^2 \vec{\rho}_\mu \vec{\rho}^\mu\right)\left(g_\omega^2 \omega_\mu \omega^\mu\right),
\end{aligned}
\end{aligned}
\end{equation}
where $\omega_{\mu\nu}=\partial_\mu \omega_\nu-\partial_\nu\omega_\mu, \vec{\rho}_{\mu \nu}=\partial_\mu \vec{\rho}_\nu-\partial_\nu \vec{\rho}_\mu,F_{\mu \nu}=\partial_\mu A_\nu-\partial_\nu A_\mu$ are the field tensors of the $\omega$, $\rho$ and photons, respectively. The isospin Pauli matrices are denoted by $\vec{\tau}$, with the third component $\tau_3 = 1$ for protons and $\tau_3 = -1$ for neutrons. $m=939~\mathrm{MeV}$ is the nucleon bare mass. $m_{\sigma}$, $m_{\omega}$ and $m_{\rho}$ are the corresponding meson masses. 
The couplings of nucleons to the mesons are denoted by $g_{\sigma}$, $g_{\omega}$, and $g_{\rho}$, while $b_{\sigma}$, $c_{\sigma}$, and $c_{\omega}$ characterize the nonlinear self-interactions of the $\sigma$ and $\omega$ mesons. The parameter $\Lambda_{\rm V}$ governs the strength of the $\rho$--$\omega$ cross-coupling. All coupling parameters are taken to be constants, independent of the baryon density.

\subsection{Density-dependent meson-exchange RMF model}
In the density-dependent meson-exchange (DDME) RMF model, instead of introducing nonlinear meson self-interactions or meson–meson couplings, the nucleon–meson couplings are assumed to be explicitly dependent on the baryon density\cite{Typel:1999yq}. The Lagrangian density in this model takes the form:
\begin{equation}\label{lagDDME}
\begin{aligned}
\mathcal{L}_{\mathrm{DD}}=  &\bar{\psi}\left(i\gamma_\mu \partial^\mu -m\right) \psi-e \bar{\psi} \gamma_\mu \frac{1+\tau_3}{2} A^\mu \psi-\frac{1}{4} F^{\mu \nu} F_{\mu \nu} \\
& +\Gamma_\sigma \sigma\bar{\psi} \psi -\Gamma_\omega \omega_\mu \bar{\psi}\gamma^\mu \psi - \Gamma_\rho \vec{\rho}_\mu \bar{\psi}\gamma^{\mu}\vec{\tau} \psi \\
&+\frac{1}{2}\left(\partial_\mu \sigma \partial^\mu \sigma-m_{\sigma}^2 \sigma^2\right)  -\frac{1}{4} \omega_{\mu \nu} \omega^{\mu \nu}+\frac{1}{2} m_\omega^2 \omega_\mu \omega^\mu \\
& -\frac{1}{4} \vec{\rho}_{\mu \nu} \vec{\rho}^{\mu \nu}+\frac{1}{2} m_\rho^2 \vec{\rho}_\mu \vec{\rho}^\mu ,
\end{aligned}
\end{equation}
where $\Gamma_{\sigma}$, $\Gamma_{\omega}$, and $\Gamma_{\rho}$ represent the density-dependent couplings, while other notations follow those of the NL RMF model. These couplings are usually parameterized as
\begin{equation}
\begin{aligned}
\Gamma_{\phi}(\rho)= & \Gamma_{\phi}\left(\rho_{0}\right) h_{\phi}(x), \quad x=\rho / \rho_0, \\
\end{aligned}
\end{equation}
with
\begin{equation}
\begin{aligned}
h_\rho(x)= & \exp \left[-a_\rho(x-1)\right] \\
h_{\phi}(x)= & a_{\phi} \frac{1+b_{\phi}\left(x+d_{\phi}\right)^2}{1+c_{\phi}\left(x+d_{\phi}\right)^2}, \quad \phi=\sigma, \omega\\
\end{aligned}
\end{equation}

\subsection{Density-dependent point-coupling RMF model}

In the density-dependent point-coupling (DDPC) RMF model, the effective Lagrangian includes isoscalar-scalar, isoscalar-vector, and isovector-vector four-fermion contact interactions~\cite{Niksic:2008vp}:

\begin{equation}\label{lagDDPC}
\begin{aligned}
\mathcal{L}_{\mathrm{PC}} & =\bar{\psi}\left(i\gamma_\mu \partial^\mu -m\right) \psi  -\frac{1}{4}F^{\mu \nu} F_{\mu \nu}-e \bar{\psi} \gamma_{\mu} \frac{\left(1+\tau_3\right)}{2}A^{\mu} \psi \\
& -\frac{1}{2} \alpha_{\mathrm{S}}(\bar{\psi} \psi)(\bar{\psi} \psi)-\frac{1}{2} \alpha_{\mathrm{V}}\left(\bar{\psi} \gamma^\mu \psi\right)\left(\bar{\psi} \gamma_\mu \psi\right)\\
&-\frac{1}{2} \alpha_{\mathrm{TV}}\left(\bar{\psi} \vec{\tau} \gamma^\mu \psi\right)\left(\bar{\psi} \vec{\tau} \gamma_\mu \psi\right) -\frac{1}{2} \delta_{\mathrm{S}}\left(\partial_\nu \bar{\psi} \psi\right)\left(\partial^\nu \bar{\psi} \psi\right),
\end{aligned}
\end{equation}
where $\delta_{\mathrm{S}}$ is the coupling parameters in derivative term, which accounts for leading effect of finite range interactions and is crucial for the description of nuclear density distribution. The couplings $\alpha_\mathrm{S}$, $\alpha_{\mathrm{V}}$, and $\alpha_{\mathrm{TV}}$ are density-dependent and parameterized as
\begin{equation}
\alpha_i(\rho)=a_i+\left(b_i+c_i x\right) e^{-d_i x}, \quad i=\mathrm{S},\mathrm{V}, \mathrm{TV}
\end{equation}
where $a_{\mathrm{TV}}$, $c_{\mathrm{TV}}$, $c_{\mathrm{V}}$ are usually set to zero for simplicity.
\subsection{Time-dependent relativistic Hartree-Bogoliubov theory and  quasiparticle finite amplitude method}

Within the framework of relativistic Hartree-Bogoliubov  (RHB) theory~\cite{Vretenar:2005zz,Meng:2005jv}, the ground-state of a nucleus is determined by solving the RHB equation
$$
\left(\begin{array}{cc}
h_D-m-\lambda & \Delta \\
-\Delta^* & -h_D^*+m+\lambda
\end{array}\right)\binom{U_\mu}{V_\mu}=E_\mu\binom{U_\mu}{V_\mu} ,
$$
where $h_D$ is the single-nucleon Dirac Hamiltonian, $\Delta$ is the pairing field,  $U$ and $V$ are Dirac spinors, $E_{\mu}$ is the quasiparticle energy, and the chemical potential $\lambda$ is determined by the particle-number conservation condition. The $h_D$ is usually derived from  effective Lagrangian density introduced in previous sub-sections under mean-field and no-sea approximations. 

Starting from the RHB ground-state solution, the quasiparticle random phase approximation (QRPA) can be formulated to describe collective excitations~\cite{Paar:2002gz, Liang:2013pda}. The QRPA equation is known to be equivalent to the time-dependent Hartree (-Fock) Bogoliubov equation at the small amplitude limit~\cite{ring2004nuclear}. We consider a ground-state nucleus is perturbed by a weak harmonic oscillating external field of frequency $\omega$:
\begin{eqnarray}
F(t)= \eta\{F(\omega)e^{-i\omega t}+F^{\dagger}(\omega)e^{+i\omega t}\}
\end{eqnarray}
with $\eta$ being a small real parameter.
In the Bogoliubov quasiparticle basis $\alpha_{\mu}$, $\alpha_{\mu}^{\dagger}$, neglecting a constant term, the external field operator $F(\omega)$ can be written as
\begin{equation*}
F(\omega) =\frac{1}{2} \sum_{\mu \nu} \{ F_{\mu \nu}^{20} \alpha_\mu^{\dagger} \alpha_\nu^{\dagger}+F_{\mu \nu}^{02} \alpha_\nu \alpha_\mu\}+\sum_{\mu \nu} F_{\mu \nu}^{11}(\omega)\alpha^{\dagger}_{\mu}\alpha_{\nu}.
\end{equation*}
For the linear response, the $F^{11}$ term does not contribute and is thus omitted. For the derivation of $F_{\mu \nu}^{20}$ and $F_{\mu \nu}^{02}$, we refer the reader to, e.g., the Appendix of Ref.~\cite{Avogadro:2011gd} for details.

The external field $F(t)$  induces  a density oscillation around the ground state with the same frequency $\omega$, which leads to the oscillation of {RHB} Hamiltonian $H(t)$ around the static  Hamiltonian $H_0$ by
\begin{eqnarray}
\delta H(t) & =&\eta\left\{\delta H(\omega) e^{-i \omega t}+\delta H^{\dagger}(\omega) e^{i \omega t}\right\}, \\
\delta H(\omega) & =&\frac{1}{2} \sum_{{\mu \nu}}\left\{\delta H_{\mu \nu}^{20}(\omega) \alpha_\mu^{\dagger} \alpha_\nu^{\dagger}+\delta H_{\mu \nu}^{02}(\omega)\alpha_\mu \alpha_\nu\right\}.
\end{eqnarray}
Similarly, due to the perturbation of $F(t)$,  the time-dependent quasiparticle  operators turns to be 
\begin{equation}
\alpha_{\mu}(t) = (\alpha_{\mu}+\delta \alpha_{\mu}(t))e^{iE_{\mu}t}.
\end{equation}
The oscillation term $\delta \alpha_{\mu}(t)$ can be expanded in
terms of quasiparticle creation operators: 
\begin{equation}
\delta \alpha_\mu(t)=\eta \sum_\nu \alpha_\nu^{\dagger}\left(X_{\nu \mu}(\omega) e^{-i \omega t}+Y_{\nu \mu}^*(\omega) e^{+i \omega t}\right) ,
\end{equation} 
where $X$ and $Y$ are forward and backward amplitudes.
 
By keeping only linear terms in $\eta$, the time-evolution equation $i\partial_t \alpha_{\mu}=[H(t)+F(t), \alpha_{\mu}(t)]$ deduces to the following linear response equations 
\begin{eqnarray}
\left(E_\mu+E_\nu-\omega\right) X_{\mu \nu}(\omega)+\delta H_{\mu \nu}^{20}(\omega)&=&-F_{\mu \nu}^{20}, \label{Eq:FAM1} \\
\left(E_\mu+E_\nu+\omega\right) Y_{\mu \nu}(\omega)+\delta H_{\mu \nu}^{02}(\omega)&=&-F_{\mu \nu}^{02}. \label{Eq:FAM2}
\end{eqnarray}
Further expanding $\delta H_{\mu \nu}^{20}(\omega)$ and $\delta H_{\mu \nu}^{02}(\omega)$ in terms of 
amplitudes  $X_{\mu \nu}(\omega)$ and $ Y_{\mu \nu}(\omega)$ up to linear orders and neglecting the right hand side results in the QRPA equations~\cite{ring2004nuclear,Hinohara:2013qda}.
The conventional QRPA method  requires calculating the QRPA matrix elements, which could be computationally expensive when the number of two-quasiparicle excitation configuration becomes very large, such as the case of deformed nuclei~\cite{Nakatsukasa:2007qj,Hinohara:2013qda}. Instead, the QFAM solves Eqs.~(\ref{Eq:FAM1}) and (\ref{Eq:FAM2}) by calculating the induced 
field $\delta H^{20}(\omega)$ and $\delta H^{02}(\omega)$ using either finite difference~\cite{Nakatsukasa:2007qj,Avogadro:2011gd,Liang:2013pda,Stoitsov:2011zz} or direct variation~\cite{Oishi:2015lph,Kortelainen:2015gxa,Sun:2017ghn,Bjelcic:2020waq}. 
Due to its numerical efficiency, the (Q)FAM has been extensively used for (Q)RPA calculations based on nuclear energy density functionals~\cite{Nakatsukasa:2007qj,Avogadro:2011gd,Oishi:2015lph,Sun:2022gdu}.

Using the solution $X$ and $Y$ from the QFAM, the strength function can be calculated as 
\begin{equation}
\frac{dB(F,\omega)}{d\omega} =-\frac{1}{\pi} \text{Im} S(F,\omega),
\end{equation}
where $S(F,\omega)$ can be expressed as 
\begin{equation}
S(F, \omega)=\frac{1}{2} \sum_{\mu \nu}\left\{F_{\mu \nu}^{20 *} X_{\mu \nu}(\omega)+F_{\mu \nu}^{02 *} Y_{\mu \nu}(\omega)\right\}
\end{equation}
in quasiparticle space. In practice, to avoid the divergence of QFAM solutions at the QRPA poles, a small imaginary part is added to the frequency $\omega \longrightarrow \omega +i\gamma$, which is equivalent to smearing the discrete RPA states with a  Lorentzian function of  width $2\gamma$~\cite{Avogadro:2011gd}.

\begin{figure}[tbp]
  \includegraphics[width=0.45\textwidth]{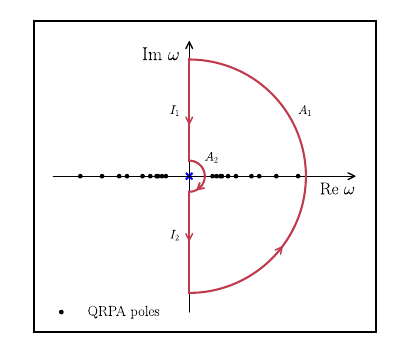}
  \caption{The contour integration path $\mathcal{C}$ (oriented counterclockwise) for the evaluation of inverse EWSR $m_{-1}$, including two semicircles $A_1$ and $A_2$, and two segments $I_1$ and $I_2$ along the vertical axis. The figure is adapted from Ref.~\cite{Hinohara:2015qra} \label{fig:contour}}
  \end{figure}

\subsection{Calculation of electric dipole polarizability}

For the electric dipole response of a spherical nucleus, we take the external field operator to be
\begin{equation}
F=\frac{N Z}{A}\left[\frac{1}{Z} \sum_{i=1}^Z r_i Y_{10}(\hat{\bm{r}}_i)-\frac{1}{N} \sum_{i=1}^N r_i Y_{1 0}(\hat{\bm{r}}_i)\right],
\end{equation}
where $A$, $Z$, $N$  are the mass, proton and neutron numbers, respectively, $r_i$ is the radial coordinate 
of the $i$-th proton or neutron, $Y_{1 0}(\hat{\bm{r}}_i)$ is the corresponding spherical harmonic function.
The electric dipole polarizability can then be evaluated using the inverse energy weighted sum rule (EWSR) $m_{-1}$
$$
\alpha_D=\frac{8 \pi e^2}{3} m_{-1}, \quad m_{-1}= \int_0^{+\infty} \frac{1}{\omega} \frac{d B(F, \omega)}{d \omega} d \omega.
$$
Instead of calculating  $m_{-1}$ directly from the strength function, we employ a 
more efficient method based on  contour integration in the complex $\omega$ plane. As proposed in Ref.~\cite{Hinohara:2015qra}, sum rules can be evaluated by using integration along a closed contour $\mathcal{C}$, sketched in Fig.\ref{fig:contour}, which includes two semicircles $A_1$ and $A_2$ centered at zero energy and two straight segments $I_1$ and $I_2$ on the imaginary axis. In particular, the $m_{-1}$ can be obtained by integrating only along the semicircle $A_2$ (clockwise) of radius $R_{A_2}$~\cite{Hinohara:2015qra} : 
\begin{equation}
m_{-1}(F)=\frac{1}{2 \pi i} \int_{{A_2}} \omega^{-1} S(F, \omega) d \omega,
\end{equation}
which accounts for the contributions from all QRPA modes with excitation energy larger than $R_{{A_2}}$. In this work, the $R_{A_2}$ is set to  $1$ MeV, and the integral is evaluated  using  Simpson's rule with 26 quadrature points.

\section{RESULTS AND DISCUSSIONS\label{Sec:result}}
Benefiting from recent high-precision experiments,  the electric dipole polarizabilities of $^{40}$Ca~\cite{Fearick:2023lyz},  $^{48}$Ca~\cite{Birkhan:2016qkr}, $^{58}$Ni~\cite{Brandherm:2024rci}, $^{68}$Ni~\cite{Rossi:2013xha}, $^{112,114,116,118,120,124}$Sn~\cite{Bassauer:2020iwp,Hashimoto:2015ema} and $^{208}$Pb~\cite{Tamii:2011pv} have been accurately measured.  However, Refs.~\cite{Bassauer:2020iwp} and~\cite{Hashimoto:2015ema} have reported considerably  different  experimental values of $\alpha_{\rm{D}}$  in $^{120}$Sn. On the theoretical side, it is known that the $\alpha_{\rm{D}}$ of open-shell Sn isotopes are significantly affected by the pairing correlations~\cite{Roca-Maza:2015eza, Li:2021aij}, leading to additional model dependence stemming from the specific choice of pairing interaction. Given these ambiguities, in the present work we restrict our focus  to the $\alpha_{\rm{D}}$ in the neutron-rich doubly (semi-)magic nuclei $^{48}\mathrm{Ca}$, $^{68}\mathrm{Ni}$, and $^{208}\mathrm{Pb}$.

To examine the sensitivity of the electric dipole polarizability to the symmetry energy and the neutron matter EOS, we consider a large set of representative nuclear EDFs that have been carefully calibrated using properties of finite nuclei,  nuclear matter, and/or neutron stars. The selected relativistic EDFs include 16 nonlinear meson-exchange models (i.e., FSU family, NL3 family, IOPB-1, IU-FSU, $\mathrm{IU}$-$\mathrm{FSU}^{*}$, FSU-Garnet, and BigApple), 10 density-dependent meson-exchange  models (i.e., DD, DD2, PK-DD, TW-99, DDME1, DDME2 and  DDMEJ family) and 13 density-dependent point-coupling RMF models (i.e., DD-PCX, DDPC-PREX, DDPC-CREX, and DDPC-REX, DD-PC1, and  DD-PCJ family). For further details on these EDFs, we refer the reader to Refs.~\cite{Sun:2023xkg,Vretenar:2003qm, Yuksel:2019dnp, Yuksel:2022umn} and references therein.
Based on these relativistic EDFs, 
the electric dipole polarizability is calculated by using the DIRQFM program~\cite{Bjelcic:2020waq,Bjelcic:2023waq}. While the original DIRQFM program supports DDME and DDPC models, we have extended it to include NL models. In
DIRQFM, the induced field $\delta H_{\mu \nu}^{20}$ and $\delta H_{\mu \nu}^{02}$ in Eqs.~(\ref{Eq:FAM1}) and (\ref{Eq:FAM2}) are calculated via direct variation. The key to incorporating the NL model lies in the modification of the induced single-nucleon Dirac Hamiltonian $\delta h_{\rm{D}}$. 
Detailed expressions for $\delta h_{\rm{D}}$ are provided in \appref. The RHB-FAM calculations are performed with 50 oscillator shells for meson fields and 22 shells for nucleon spinors after convergence check.  For the pairing field, we employ a separable pairing interaction reproducing pairing gap of the D1S Gogny force in symmetric nuclear matter (see Ref.~\cite{Niksic:2014dra} for details).
In addition to the relativistic EDFs, we also consider 24 nonrelativistic Skyrme EDFs (namely, SIII, SIV, SV, SVI, SLy230a, SLy230b, SLy4, SLy5, SLy8, SAMi, SAMi-J30, SAMi-J31, SAMi-J32, SAMi-J33, SGI, SGII, SkM, SkM*, Ska, MSk1, MSk2, MSk7, BSk1, and BSk2). The predicted values of $\alpha_{\rm{D}}$ in $^{48}$Ca, $^{68}$Ni, and $^{208}$Pb, obtained from QRPA calculations with these EDFs, are adopted from Ref.~\cite{Li:2021aij} and used in our analysis.

\begin{figure*}[htbp]
  \includegraphics[width=1\linewidth]{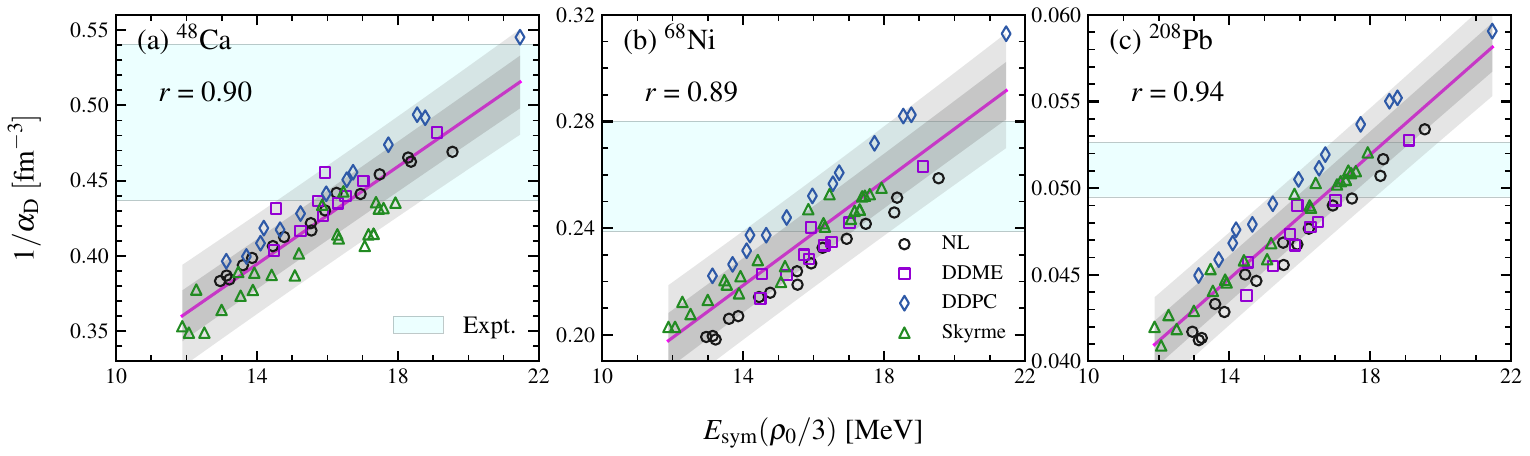}
  \caption{Inverse dipole polarizabilities of $^{48}\mathrm{Ca}$ (a), $^{68}\mathrm{Ni}$ (b), and $^{208}\mathrm{Pb}$ (c) as functions of the symmetry energy at density $\rho_0/3$, calculated with various nuclear energy density functionals. Results from the NL, DDME, DDPC, and Skyrme models are represented by open circles, squares, diamonds, and upward triangles, respectively.  Pearson correlation coefficients $r=0.90$, $0.89$ and $0.94$ respectively for the three nuclei are displayed. The linear fits to the predictions are shown as the solid line and the $68\%$($95\%)$ prediction bands are displayed by the darker (lighter) shadowed regions. For comparison, experimental data from Refs.~\cite{Tamii:2011pv, Rossi:2013xha, Birkhan:2016qkr} are indicated by the light cyan bands\label{fig:ad-esym}.}
  \end{figure*}

\subsection{Symmetry energy and electric dipole polarizability\label{sec:sym}}

Figures~\ref{fig:ad-esym}(a), (b), and (c) show the inverse dipole polarizabilities $1/\alpha_{\rm{D}}$ of $^{48}\mathrm{Ca}$, $^{68}\mathrm{Ni}$, and $^{208}\mathrm{Pb}$ respectively, as functions of the symmetry energy at  $\rho_0/3$,  $E_{\rm{sym}}(\rho_0/3)$. The predictions are obtained from NL (open circles), DDME (open squares), DDPC (open diamonds), and Skyrme (open upward triangles) EDFs. For each nucleus, a  linear fit is performed using all 63 EDFs, shown as solid lines in the corresponding windows. The inner and outer shaded bands indicate the $68\%$ and $90\%$ prediction intervals, respectively. For comparison, the experimentally measured values of $\alpha_{\rm{D}}$ corrected for quasi-deuteron contributions~\cite{Roca-Maza:2015eza} are also shown: $\alpha_{\mathrm{D}}^{48}=2.07\pm 0.22~\mathrm{fm}^3$~\cite{Birkhan:2016qkr}, $\alpha_{\mathrm{D}}^{68}=3.88\pm 0.31\mathrm{fm}^3$~\cite{Rossi:2013xha}, and $\alpha_{\mathrm{D}}^{208}=19.6\pm 0.6\mathrm{fm}^3$~\cite{Tamii:2011pv}.

It is seen from Fig.~\ref{fig:ad-esym}, within each class of nuclear EDFs, there exist strong linear correlations between $1/\alpha_{\mathrm{D}}$ and $E_{\mathrm{sym}}(\rho_0/3)$ for all the three nuclei. 
The observed sensitivity of $1/\alpha_{\mathrm{D}}$ to the symmetry energy at $\rho_0/3$ can be understood by the significant role of the surface symmetry energy on $\alpha_{\rm{D}}$, 
which has been revealed by early studies based on the macroscopic hydrodynamic  and droplet models (see, e.g., Ref.~\cite{Zhang:2015ava}). However, when comparing  predictions of different types of models, one clearly sees noticeable systematic uncertainties. For instance, the DDPC models tend to systematically predict higher values of $1/\alpha_{\rm{D}}$ for $^{208}$Pb compared to other three model families. Such model dependence may stem from difference in the intrinsic density-dependence in various types of nuclear EDFs and requires further studies. 

Quantitatively, the linear fits yield the following relations:
\begin{eqnarray}
\frac{10^{3}}{\alpha_{\rm{D}}^{48}} &=& (168\pm16)+(16.2\pm 1.0)  E_{\mathrm{sym}}(\rho_0/3), \label{Eq:Ad48}\\ 
\frac{10^{3}}{\alpha_{\rm{D}}^{68}} &=& (81.5\pm9.9)+(9.79\pm 0.63) E_{\mathrm{sym}}(\rho_0/3), \label{Eq:Ad68}\\ 
\frac{10^{3}}{\alpha_{\rm{D}}^{208}} &=& (1.80\pm 0.08)+(19.7\pm 1.3)E_{\mathrm{sym}}(\rho_0/3), \label{Eq:Ad208}
\end{eqnarray}
where $\alpha_{\rm{D}}$ and $E_{\rm{sym}}(\rho_0/3)$ are in units of $\rm{fm}^3$ and MeV, respectively. 
Using these linear relations, the predicted mean values of $1/\alpha_{\mathrm{D}}$ in $^{48}$Ca, $^{68}$Ni and $^{208}$Pb at a given  $E_{\mathrm{sym}}(\rho_0/3)$ (denoted by $x^*$) can be evaluated, along with the associated $1\sigma$ prediction uncertainties given by~\cite{Dra1998}:
\begin{equation}
\sigma^{\mathrm{inv}} = s \cdot \left\{1 + \frac{1}{n} + \frac{(x^* - \bar{x})^2}{\sum (x_i - \bar{x})^2} \right\}^{1/2}, \label{eq:sigma_inv}
\end{equation}
where $n$ is the number of EDFs, and the unbiased residual variance  is defined as $s^2 = \frac{1}{n - 2} \sum_{i=1}^n (y_i - \hat{y}_i)^2$. Here, $(x_i, y_i)$ denote the calculated values of $[ E_{\mathrm{sym}}(\rho_0/3),1/\alpha_{\mathrm{D}}]$ using the $i$-th EDF, $\bar{x}$ is the averaged value of all $x_i$, and $\hat{y}_i$ is the corresponding value at $x_i$ from linear regression.

\begin{figure}[hbtp]
  \includegraphics[width=1\linewidth]{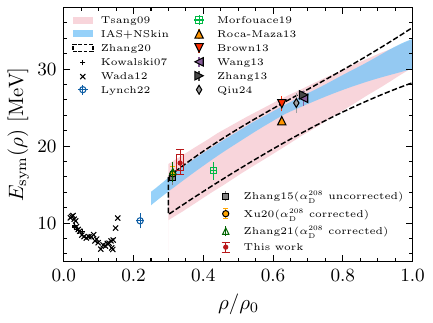}
  \caption{ Constraints on the symmetry energy $E_{\rm{sym}}(\rho)$ as a function of density $\rho$ (see text for the details). The inferred $E_{\mathrm{sym}}(\rho_0/3)$ obtained in this work is displayed as the solid circle, along with  the 68$\%$ and $90\%$ confidence intervals are depcited by  the  rectangle and  error bar, respectively.  \label{fig:EsymCon}}
  \end{figure}
  
 \begin{figure*}[!bhtp]
  \includegraphics[width=1\linewidth]{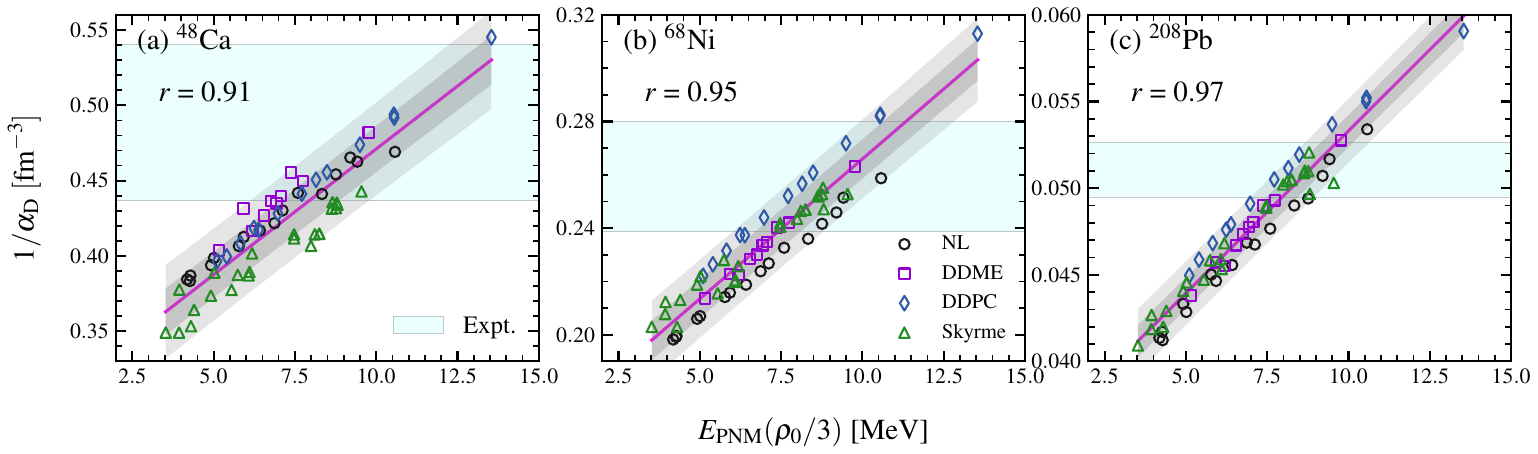}
  \caption{Same as Fig.~\ref{fig:ad-esym}, but for the neutron matter EOS at $\rho_0/3$\label{fig:ad-epnm}.}
  \end{figure*}
  
Using the obtained linear relations, we perform Bayesian inference to constrain $E_{\mathrm{sym}}(\rho_0/3)$ based on the measured electric dipole polarizabilities of $^{48}\mathrm{Ca}$, $^{68}\mathrm{Ni}$, and $^{208}\mathrm{Pb}$. Given the experimental data $\bm{\mathcal{O}}^{\mathrm{exp}} = (\alpha_{\mathrm{D}}^{48}, \alpha_{\mathrm{D}}^{68}, \alpha_{\mathrm{D}}^{208})$, the posterior distribution of $\theta \equiv E_{\mathrm{sym}}(\rho_0/3)$ is determined by Bayes' rule
\begin{equation}
p(\theta \mid \bm{\mathcal{O}}^{\mathrm{exp}}) \propto p(\bm{\mathcal{O}}^{\mathrm{exp}} \mid \theta)\, \pi(\theta), \label{eq:posterior}
\end{equation}
where a uniform prior $\pi(\theta)$ is adopted over the range 0--30 MeV. The likelihood function is assumed to be Gaussian:
\begin{equation}
p(\bm{\mathcal{O}}^{\mathrm{exp}} \mid \theta) \propto \exp\left[-\sum_i \frac{(\mathcal{O}_i(\theta) - \mathcal{O}_i^{\mathrm{exp}})^2}{2 \sigma_i^2}\right],
\end{equation}
where $\mathcal{O}_i(\theta)$ are the predicted $\alpha_{\mathrm{D}}$ values from the linear relations Eqs. (\ref{Eq:Ad48}-\ref{Eq:Ad208}). The total uncertainty $\sigma_i$ includes both experimental and theoretical errors:
\[
\sigma_i^2 = (\sigma_i^{\mathrm{exp}})^2 + \left[\frac{\sigma_i^{\mathrm{inv}}}{\mathcal{O}_i^2(\theta)}\right]^2.
\]
Using the sequential Monte Carlo algorithm~\cite{abril2023pymc} with 10 chains of $10^5$ samples each, we infer from the Bayesian analysis
\begin{equation}
E_{\mathrm{sym}}(\rho_0/3) = 17.8^{+1.1(1.8)}_{-0.9(1.6)}~\mathrm{MeV},
\end{equation}
at 68\% (90\%) confidence level.

Figure~\ref{fig:EsymCon} shows the inferred $E_{\mathrm{sym}}(\rho_0/3)$ as a red solid circle, with the 68\% confidence interval shown as a rectangle and the 90\% interval illustrated by capped error bar. 
For comparison, we also show in Fig.~\ref{fig:EsymCon} constraints on the symmetry energy at subsaturation densities from other analyses of nuclear structure probes and observables in heavy-ion collisions. Throughout, the saturation density is assumed to be $0.16~\mathrm{fm}^{-3}$. The constraints from the Skyrme-Hartree-Fock  analyses 
of isobaric analogue states and neutron skin thickness data are shown as the blue shaded band (IAS+NSkin)~\cite{Danielewicz2014}, while 
those from two transport model analyses of isospin diffusion are presented as the pink shaded band (Tsang09)~\cite{Tsang:2008fd} and  black dashed region (Zhang20)~\cite{Zhang:2020azr}, respectively.
At densities below $0.2\rho_0$, we show  the  results for the symmetry energies at finite temperature of $3\sim11$ MeV extracted from cluster formation in heavy-ion collisions
(Wada12 and Kowalski07)~\cite{Wada:2011qm, Kowalski:2006ju}.
At higher densities, symbols in Fig.~\ref{fig:EsymCon} indicate
a constraint on the $E_{\rm{sym}}(\rho)$ around $0.21\rho_0$ from the isospin diffusion (Lynch22)~\cite{Lynch:2021xkq}, a constraint around $0.43 \rho_0$ from Bayesian analysis of single and double neutron-proton ratio (Morfouace19)~\cite{ Morfouace:2019jky}, and five constraints around $2\rho_0/3$ from the giant quadrupole resonance in $^{208}$Pb (Roca-Maza13)~\cite{Roca-Maza:2012uor}, 
properties of doubly magic nuclei
(Brown13)~\cite{Brown:2013mga}, the effective proton-neutron chemical potential difference (Qiu24)~\cite{Qiu:2023kfu},
the neutron-proton Fermi-energy difference  (Wang13)~\cite{Wang:2013yra}, and the binding energy difference between heavy isotope pairs (Zhang13)~\cite{Zhang:2013wna}.

In particular, three constraints on $E_{\mathrm{sym}}(0.05~\mathrm{fm}^{-3})$ extracted from the $\alpha_{\rm{D}}$ in $^{208}$Pb are also included in Fig.~\ref{fig:EsymCon}. We note that the constraint $E_{\mathrm{sym}}(0.05~\mathrm{fm}^{-3})= 15.91 \pm 0.99$ MeV  reported in~\cite{Zhang:2015ava} (Zhang15) was extracted from an earlier  experimental value of $\alpha^{208}_{\rm{D}}= 20.1\pm0.6~\rm{fm}^3$ by Tamii \textit{et al.}~\cite{Tamii:2011pv}. As pointed out by Roca-Maza \textit{et al.}~\cite{Roca-Maza:2015eza}, this value includes the contributions from quasideuteron excitations, which should be subtracted for comparison with theoretical (Q)RPA calculation.
Using the corrected value $\alpha^{208}_{\rm{D}} = 19.6\pm0.6~\rm{fm}^3$, Bayesian analyses
based on the Skyrme EDF in Refs.~\cite{Xu:2020xib} and~\cite{Zhang:2021vuj} obtain larger values of
$E_{\rm{sym}}(0.05~\mathrm{fm}^{-3}) = 16.4^{+1.0}_{-0.9}$ MeV (Xu20)
and $E_{\rm{sym}}(0.05~\mathrm{fm}^{-3}) = 16.7\pm{1.3}$ MeV (Zhang21), respectively. In this work, by employing the corrected experimental value and additionally including data  on the $\alpha_{\rm{D}}$ of 
$^{48}$Ca and $^{68}$Ni, we obtain a relatively large value of $E_{\rm{sym}}(\rho_0/3)$ compared to other constraints. Nevertheless, our result remains consistent with the “IAS+NSkin” constraint and those from isospin diffusion data within the $1\sigma$ uncertainty band.

\subsection{Neutron matter equation of state and electric dipole polarizability}

Under the empirical parabolic approximation, the EOS  of pure neutron matter  can be expressed as
\begin{equation}
\begin{aligned}
E_{\mathrm{PNM}}(\rho)\approx E_{0}(\rho)+E_{\mathrm{sym}}(\rho),
\end{aligned}
\end{equation}
where  $E_0(\rho)$ is the EOS of  symmetric nuclear matter. Since the $E_{0}(\rho)$ at subsaturation densities is relatively well determined by properties of finite nuclei, one expects the electric dipole polarizability to be sensitive to $E_{\mathrm{PNM}}(\rho_0/3)$.

 In Fig.~\ref{fig:ad-epnm}, we present the inverse electric dipole polarizabilities $1/\alpha_{\mathrm{D}}$ of $^{48}\mathrm{Ca}$, $^{68}\mathrm{Ni}$, and $^{208}\mathrm{Pb}$ as functions of the pure neutron matter EOS $E_{\mathrm{PNM}}(\rho_0/3)$ predicted by four classes of nuclear energy density functionals. Similar to the correlations with $E_{\mathrm{sym}}(\rho_0/3)$ shown in Fig.~\ref{fig:ad-esym}, strong linear relationships are observed between $1/\alpha_{\mathrm{D}}$ and $E_{\mathrm{PNM}}(\rho_0/3)$ for all three nuclei. However, the correlations become weaker for lighter nuclei, accompanied by increased systematic discrepancies among different types of models. Notably, compared to the $1/\alpha_{\mathrm{D}}$--$E_{\mathrm{sym}}(\rho_0/3)$ relations, the $1/\alpha_{\mathrm{D}}$--$E_{\mathrm{PNM}}(\rho_0/3)$ correlations exhibit weaker model dependence, particularly for $^{208}$Pb.

\begin{figure}[hbtp]
  \includegraphics[width=1\linewidth]{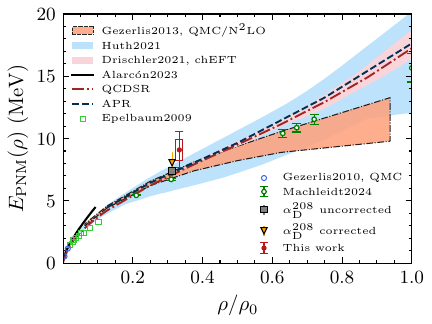}
  \caption{Constraints and predictions on the pure neutron matter EOS $E_{\rm{PNM}}(\rho)$ at  subsaturation densities (see text for the details). The inferred value of $E_{\mathrm{PNM}}(\rho_0/3)$ from this work is shown as a solid circle. The associated $68\%$ and $90\%$ confidence intervals are indicated by the rectangle and the error bar, respectively.\label{fig:EpnmCon}}
  \end{figure}

Following a similar approach as in Sec.~\ref{sec:sym}, we conduct linear regressions for the $1/\alpha_{\mathrm{D}}$--$E_{\mathrm{PNM}}(\rho_0/3)$ relations predicted by all the EDFs, and obtain 
\begin{eqnarray}
\frac{10^{3}}{\alpha_{\rm{D}}^{48}} &=& (304\pm 7) +(16.7\pm 1.0)E_{\mathrm{PNM}}(\rho_0/3),\\ 
\frac{10^{3}}{\alpha_{\rm{D}}^{68}} &=& (161\pm 3)+(10.5\pm 0.5)E_{\mathrm{PNM}}(\rho_0/3),\\ 
\frac{10^{3}}{\alpha_{\rm{D}}^{208}} &=& (34.6\pm 0.4 )+(1.87\pm 0.06)E_{\mathrm{PNM}}(\rho_0/3).
\end{eqnarray}
Using these linear relations, we  extract $E_{\mathrm{PNM}}(\rho_0/3)$  from Bayesian analysis using the experimental data on 
the $\alpha_{\rm{D}}$ in $^{48}$Ca, $^{68}$Ni and $^{208}$Pb. A uniform prior in the range of 0--30 MeV is adopted for the analysis. The resulting constraint is 
\begin{equation}
E_{\mathrm{PNM}}(\rho_0/3)= 9.1^{+0.8(1.4)}_{-0.9(1.4)}~\mathrm{MeV}
\end{equation}
at a 68(90)$\%$ confidence level. This result is shown as  the red solid circle in Fig.~\ref{fig:EpnmCon}  with the $68\%$ and $90\%$ confidence intervals indicated by a rectangle and an capped error bar, respectively.

For comparison, Fig.~\ref{fig:EpnmCon} also shows predictions on the $E_{\rm{PNM}}(\rho)$ at subsaturation densities from 
auxiliary-field diffusion Monte Carlo calculations based on local next-to-next-to-leading order ($\mathrm{N}^2$LO) NN interactions (Gezerlis2013, QMC/$\rm{N}^2$LO)~\cite{Gezerlis:2013ipa},
chiral effective field theory  (Huth2021)~\cite{Huth:2020ozf},
many-body perturbation theory calculations with NN and 3N chiral interactions up to fourth order (N$^3$LO) by Drischler \textit{et al.} (Drischler2021, ChEFT)~\cite{Drischler:2020hwi}, ladder resummation (Alarcón2023)~\cite{Alarcon:2022vtn}, QCD sum rules (QCDSR)~\cite{Cai:2019vsg}
and  variational calculations by Akmal-Pandharipande-Ravenhall (APR)~\cite{Akmal:1998cf}, next-to-leading order lattice calculation (Epelbaum2009)~\cite{Epelbaum:2009rkz},  quantum Monte-Carlo simulations for low density neutron matter (Gezerlis2010)~\cite{Gezerlis:2009iw}, and N$^3$LO chiral EFT predictions by Machleidt and Sammarruca (Machleidt2024)~\cite{Machleidt:2024bwl}. In addition, the constraint $E_{\rm{PNM}}(0.05~\mathrm{fm}^{-3}) = 7.39 \pm 0.81~\mathrm{MeV}$ ($E_{\rm{PNM}}(0.05~\mathrm{fm}^{-3}) = 8.06 \pm 0.85~\mathrm{MeV}$)~\cite{Zhang:2015ava}, extracted from the uncorrected (corrected) experimental value of $\alpha_{\mathrm{D}}^{208}$, is shown as a solid square (downward triangle) in Fig~\ref{fig:EpnmCon}. Unless explicitly stated in the original reference, we adopt a default saturation density of $\rho_0 = 0.16~\mathrm{fm}^{-3}$. Notable exceptions include $\rho_0 = 0.17~\mathrm{fm}^{-3}$ in the chiral EFT results by Drischler \textit{et al.}\cite{Drischler:2020hwi}, and $\rho_0 = 0.155~\mathrm{fm}^{-3}$ in the N$^3$LO chiral EFT predictions by Machleidt and Sammarruca~\cite{Machleidt:2024bwl}.

As can be seen in Fig.~\ref{fig:EpnmCon}, the constraint on $E_{\rm{PNM}}(\rho_0/3)$ obtained in the present work lies above the predictions from microscopic calculations. Notably, even the lower bound of the resulting $1\sigma$ confidence interval is at the upper edge of the uncertainty band derived from various chiral effective field theory predictions, suggesting a mild  tension between the nuclear EDF-based analysis of $\alpha_{\rm D}$ data and microscopic many-body calculations.

\begin{figure*}[t]
  \includegraphics[width=0.66\linewidth]{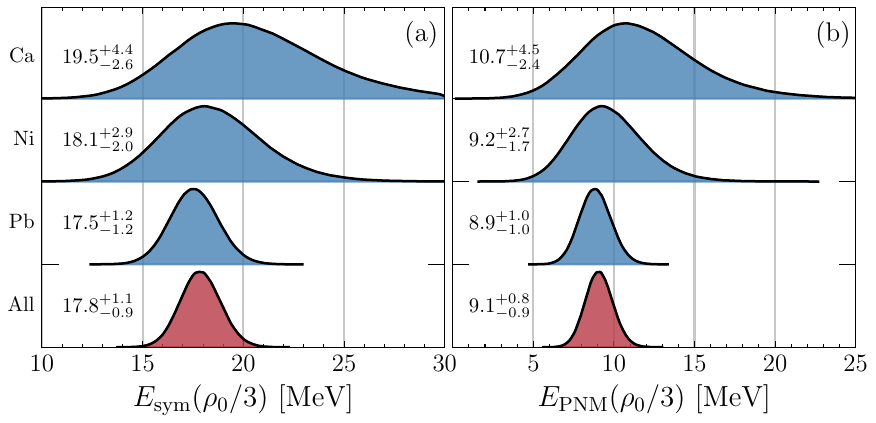}
  \caption{Posterior distribution function of symmetry energy and neutron matter EOS at $\rho_0/3$  under the respective constraints of $^{48}\mathrm{Ca}$, $^{68}\mathrm{Ni}$, 
  $^{208}\mathrm{Pb}$ and their joint analyses (labelled by `All').\label{fig:PDF}}
  \end{figure*}

Before closing this section, we show in Fig.~\ref{fig:PDF}(a) and (b) the posterior distributions of $E_{\rm{sym}}(\rho_0/3)$ and $E_{\rm{PNM}}(\rho_0/3)$, respectively, as inferred from Bayesian analyses based on the $\alpha_{\rm{D}}$ of each nucleus individually, as well as from the combined analysis (denoted as “All”). The corresponding inferred maximum a posterior values and their $1\sigma$ confidence intervals are also indicated. It is seen that the experimental values of $\alpha_{\rm{D}}$ in $^{48}$Ca and $^{68}$Ni favor larger 
$E_{\rm{sym}}(\rho_0/3)$ and $E_{\rm{PNM}}(\rho_0/3)$ compared to those inferred from $^{208}$Pb. Nevertheless, the results from the combined analysis (“All”) closely follow those inferred from $^{208}$Pb alone, indicating that the constraints on $E_{\rm{sym}}(\rho_0/3)$ and $E_{\rm{PNM}}(\rho_0/3)$ are dominated by the $\alpha_{\rm{D}}$ data of $^{208}$Pb. This is understandable, given that both the experimental uncertainty in $1/\alpha_{\rm{D}}$ and the intra- and inter-model uncertainties in the $1/\alpha_{\rm{D}}$--$E_{\rm{sym/PNM}}$ correlations are significantly smaller for $^{208}$Pb compared to the lighter nuclei.

\section{Conclusion\label{Sec:conclusion}}
By employing a number of representative parameter sets for three classes of relativistic nuclear energy density functionals (i.e.,  nonlinear meson-exchange model, density-dependent meson-exchange model and density-dependent point-coupling model) and the nonrelativistic Skyrme energy density functionals, we have demonstrated that the inverse electric dipole polarizabilities $1/\alpha_{\mathrm{D}}$ of $^{48}\mathrm{Ca}$, $^{68}\mathrm{Ni}$, and $^{208}\mathrm{Pb}$ exhibit strong linear correlations with both the symmetry energy $E_{\rm{sym}}(\rho)$ and the pure neutron matter equation of state $E_{\rm{PNM}}(\rho)$ at a subsaturation density $\rho = \rho_0/3$.  Remarkably, the $1/\alpha_{\mathrm{D}}$--$E_{\rm{PNM}}(\rho_0/3)$ correlation for  $^{208}$Pb shows minimal model dependence.

Using linear regression analyses of the model predictions  and then applying Bayesian analysis of  experimental  data for $\alpha_{\rm{D}}$ in $^{48}$Ca, $^{68}$Ni and $^{208}$Pb, we infer  $E_{\mathrm{sym}}(\rho_0/3) = 17.8^{+1.1(1.8)}_{-0.9(1.6)}~\mathrm{MeV}$ and $E_{\mathrm{PNM}}(\rho_0/3) = 9.1^{+0.8(1.4)}_{-0.9(1.4)}~\mathrm{MeV}$
at the 68\% (90\%) confidence level.  The extracted value of $E_{\mathrm{sym}}(\rho_0/3)$ is relatively high but remains consistent with empirical estimates within the $1\sigma$ uncertainty. In contrast, the inferred $E_{\mathrm{PNM}}(\rho_0/3)$ exceeds most predictions from microscopic many-body theories, and its $1\sigma$ interval does not overlap with the range derived from  the chiral effective field theory. This suggests a mild tension between EDF-based constraints derived from $\alpha_{\mathrm{D}}$ data and the results of microscopic calculations.

\begin{acknowledgments}
This work was supported in part by the National Natural Science Foundation of China under Grant No. 12235010, the National SKA Program of China No. 2020SKA0120300,  the Science and Technology Commission of Shanghai Municipality under Grant No. 23JC1402700, the Lingchuang Research Project of China National Nuclear Corporation under Grant No. CNNC-LCKY-2024-082, and the National Key Research and Development Program under Grants No.2021YFA1601500. ZZ would like to thank Li-Gang Cao and Jorge Piekarewicz for helpful discussions on RPA calculations based on the nonlinear RMF model. We thank Tamara Nikšić  and Ante Ravlić for providing the parameter values of the DD-MEJ family.
\end{acknowledgments}

\appendix*
\section{Induced fields for nonlinear meson exchange model\label{Sec:ind}}
The single-nucleon Dirac Hamiltonian for nonlinear meson-exchange RMF model is given by~\cite{Bjelcic:2023waq}
\begin{equation}
h_D=\left[\begin{array}{cc}
\Sigma^0+\left(\Sigma_s+m\right) & \boldsymbol{\sigma} \cdot(\boldsymbol{p}-\boldsymbol{\Sigma}), \\
\boldsymbol{\sigma} \cdot(\boldsymbol{p}-\boldsymbol{\Sigma}) & \Sigma^0-\left(\Sigma_s+m\right)
\end{array}\right] ,
\end{equation}
where $\boldsymbol{\sigma}$ are the Pauli matrices, and the scalar and vector self-energies are given by
\begin{equation}
\begin{aligned}
\Sigma_s & =-g_\sigma \sigma, \\
\Sigma^\mu & =g_\omega \omega^\mu+g_\rho \vec{\tau}  \vec{\rho}^\mu+e \frac{1+\tau_3}{2} A^\mu,\\
\end{aligned}
\end{equation}
with $\Sigma^0$ and $\boldsymbol{\Sigma}$ denoting the time-like and space-like components of $\Sigma^{\mu}$, respectively.

In the presence of external fields, the induced single-nucleon Dirac Hamiltonian are obtained via functional variations with respect to the densities
\begin{equation}
\begin{aligned}
& \delta h_{{D}}=\left[\begin{array}{cc}
\delta \Sigma^{0}+\delta \Sigma_s & -\boldsymbol{\sigma} \cdot \delta \boldsymbol{\Sigma} \\
-\boldsymbol{\sigma} \cdot \delta \boldsymbol{\Sigma} & \delta \Sigma^{0}-\delta \Sigma_s
\end{array}\right], \\
\end{aligned}
\end{equation}
where the corresponding induced self-energies are
\begin{equation}\label{Eq:hd}
\begin{aligned}
& \delta \Sigma_s= -g_\sigma \delta \sigma ,\\
& \delta \Sigma^0=g_\omega \delta \omega^0+\tau_3 g_\rho \delta \rho^0+\frac{1+\tau_3}{2} \delta V_C, \\
&\delta \boldsymbol{\Sigma}=g_\omega \delta \boldsymbol{\omega}+\tau_3 g_\rho \delta \boldsymbol{{\rho}}+\frac{1+\tau_3}{2} \delta \mathbf{V}_C.
\end{aligned}
\end{equation}
Here, $\delta V_C$ and $\delta \mathbf{V}_C$ are the time-like and space-like components of the induced Coulomb field, respectively, obtained by solving the Poisson equations
\begin{equation}
\begin{aligned}
-\nabla^2 \delta V_C=e^2 \delta \rho_p, \\
-\nabla^2 \delta \mathbf{V}_C=e^2 \delta \boldsymbol{j}_p,
\end{aligned}
\end{equation}
where $\delta \rho_p$ and $\delta \boldsymbol{j}_p$ are the induced density and spatial current of protons. The induced meson fields are governed by linearized Klein-Gordon equations. For the scalar and time-like vector meson fields, one obtains
\begin{equation}
\begin{aligned}
\left[-\Delta+m_\sigma^2\right] \delta \sigma=&g_\sigma \delta \rho_{\mathrm{S}}-2 b_\sigma m g_\sigma^3[\sigma]_{\mathrm{GS}} \delta \sigma\\
& -3 c_\sigma g_\sigma^4[\sigma]_{\mathrm{GS}}^2 \delta \sigma ,\\
\left[-\Delta+m_\omega^2\right] \delta \omega^0=&g_\omega \delta \rho_{\mathrm{B}}-3 c_\omega g_\omega^4\left[\omega^0\right]_{\mathrm{GS}}^2 \delta \omega^0\\
&-\Lambda_{\rm{V}} g_\rho^2\left[\rho^0\right]_{\mathrm{GS}}^2 g_\omega^2 \delta \omega^0\\
& -2 \Lambda_{\rm{V}} g_\rho^2 g_\omega^2\left[\rho^0 \omega^0\right]_{\mathrm{GS}} \delta \rho^0 ,\\
\left[-\Delta+m_\rho^2\right] \delta \rho^0=&g_\rho \delta \rho_{3,\mathrm{B}}-\Lambda_{\rm{V}} g_\rho^2 g_\omega^2\left[\omega^0\right]_{\mathrm{GS}}^2 \delta \rho^0\\
&-2 \Lambda_{\rm{V}} g_\rho^2 g_\omega^2\left[\rho^0 \omega^0\right]_{\mathrm{GS}} \delta \omega^0, \\
\end{aligned}
\end{equation}
and the equations for the space-like components of the vector fields read
\begin{equation}
\begin{aligned}
\left[-\Delta+m_\omega^2\right] \delta \boldsymbol{\omega}=&g_\omega \delta \boldsymbol{j}-c_\omega g_\omega^4\left[\omega^0\right]_{\mathrm{GS}}^2 \delta \boldsymbol{\omega}\\
&-\Lambda_{\rm{V}} g_\rho^2\left[\rho^0\right]_{\mathrm{GS}}^2 g_\omega^2 \delta \boldsymbol{\omega}, \\
\left[-\Delta+m_\rho^2\right] \delta \boldsymbol{\rho}=&g_\rho \delta \boldsymbol{j}_3-\Lambda_{\rm{V}} g_\rho^2 g_\omega^2\left[\omega_0\right]_{\mathrm{GS}}^2 \delta \boldsymbol{\rho}.
\end{aligned}
\end{equation}
Here, the subscript “GS” denotes the ground state static fields. 
The relevant densities and space-like currents entering the Klein-Gordon equations are defined as
\begin{equation}
\begin{aligned}
\rho_{\mathrm{S}}(\boldsymbol{r}) &= \sum_{i=1}^A \bar{\psi}_i(\boldsymbol{r}) \psi_i(\boldsymbol{r}), \\
\rho_{\mathrm{B}}(\boldsymbol{r}) &= \sum_{i=1}^A \bar{\psi}_i(\boldsymbol{r}) \gamma^0 \psi_i(\boldsymbol{r}), \\
\rho_{3,\mathrm{B}}(\boldsymbol{r}) &= \sum_{i=1}^A \bar{\psi}_i(\boldsymbol{r}) \tau_3 \gamma^0 \psi_i(\boldsymbol{r}),
\end{aligned}
\end{equation}
and
\begin{equation}
\begin{aligned}
\boldsymbol{j}(\boldsymbol{r}) &= \sum_{i=1}^A \bar{\psi}_i(\boldsymbol{r}) \boldsymbol{\gamma} \psi_i(\boldsymbol{r}), \\
\boldsymbol{j}_3(\boldsymbol{r}) &= \sum_{i=1}^A \bar{\psi}_i(\boldsymbol{r}) \tau_3 \boldsymbol{\gamma} \psi_i(\boldsymbol{r}),
\end{aligned}
\end{equation}
where the bold symbols  represent the space-like components of the corresponding four-currents, i.e., $\boldsymbol{\gamma} = (\gamma^1,\gamma^2,\gamma^3)$ .

\bibliography{apssamp}

\end{document}